\documentclass{jfm-arxiv}

\usepackage{graphicx,bm, amsmath,mathrsfs}
\usepackage{natbib}

\ifCUPmtlplainloaded \else
  \checkfont{eurm10}
  \iffontfound
    \IfFileExists{upmath.sty}
      {\typeout{^^JFound AMS Euler Roman fonts on the system,
                   using the 'upmath' package.^^J}%
       \usepackage{upmath}}
      {\typeout{^^JFound AMS Euler Roman fonts on the system, but you
                   dont seem to have the}%
       \typeout{'upmath' package installed. JFM.cls can take advantage
                 of these fonts,^^Jif you use 'upmath' package.^^J}%
       \providecommand\upi{\pi}%
      }
  \else
    \providecommand\upi{\pi}%
  \fi
\fi

\ifCUPmtlplainloaded \else
  \checkfont{msam10}
  \iffontfound
    \IfFileExists{amssymb.sty}
      {\typeout{^^JFound AMS Symbol fonts on the system, using the
                'amssymb' package.^^J}%
       \usepackage{amssymb}%
       \let\le=\leqslant  \let\leq=\leqslant
       \let\ge=\geqslant  
      }{}
  \fi
\fi

\ifCUPmtlplainloaded \else
  \IfFileExists{amsbsy.sty}
    {\typeout{^^JFound the 'amsbsy' package on the system, using it.^^J}%
     \usepackage{amsbsy}}
    {\providecommand\boldsymbol[1]{\mbox{\boldmath $##1$}}}
\fi

\providecommand\bnabla{\boldsymbol{\nabla}}
\providecommand\bcdot{\boldsymbol{\cdot}}

\newsavebox{\astrutbox}
\sbox{\astrutbox}{\rule[-5pt]{0pt}{20pt}}

\newcommand\etal{\mbox{\textit{et al.}}}

\title[Orientation of non-spherical particles in an
axisymmetric random flow]{Orientation of
non-spherical particles in an axisymmetric random flow}

\author[D. Vincenzi]{D\ls A\ls R\ls I\ls O\ns
V\ls I\ls N\ls C\ls E\ls N\ls Z\ls I}

\affiliation{CNRS UMR 7351, Laboratoire J.-A. Dieudonn\'e, Universit\'e de 
Nice Sophia Antipolis, \\  Parc Valrose, 06108 Nice, France}

\pubyear{}
\volume{}
\pagerange{}
\date{?? and in revised form ??}

\begin{document}

\maketitle

\begin{abstract}
The dynamics of non-spherical rigid particles immersed in an
axisymmetric random flow is studied analytically. 
The motion of the particles is described by Jeffery's equation;
the random 
flow is Gaussian and has short correlation time.
The stationary probability density function of orientations is calculated
exactly. Four regimes are identified depending on the statistical 
anisotropy of the flow and on the geometrical shape of the particle.
If~$\boldsymbol{\lambda}$ is the axis of symmetry of the flow,
the four regimes are: 
rotation about~$\boldsymbol{\lambda}$,
tumbling motion between~$\boldsymbol{\lambda}$ and~$-\boldsymbol{\lambda}$, 
combination of rotation and tumbling, and
preferential alignment with a direction 
oblique to~$\boldsymbol{\lambda}$.
\end{abstract}


\section{Introduction}

Non-spherical solid particles suspended in a moving fluid
rotate and orient themselves under the action of the velocity gradient. 
Even at low concentrations, the orientational dynamics of non-spherical
particles can influence the rheological properties of a suspension, namely
the intrinsic viscosity and the normal stress coefficients
\citep*{BHAC77,L99}. This phenomenon has diverse practical applications. 
In the turbulent regime, for instance,
the injection of rodlike polymers in a Newtonian fluid
can produce a considerable reduction of the turbulent drag,
with this effect being routinely exploited to reduce energy losses in  
pipelines \citep*{GB95}.
The study of the orientation of particles immersed in a fluid
also has numerous applications in the natural sciences.
Amongst them it is worth mentioning the swimming motion of
certain biological micro-organisms \citep{SS08,KS11} and the
formation of ice crystals in clouds \citep{CL94}.
This latter phenomenon plays a crucial role in
processes such as rain initiation and radiative transfer. 

The starting point for understanding the properties of a dilute 
suspension is the motion of an isolated particle in a given
flow field.
Analytical results on the dynamics of a non-spherical particle
have been obtained for various laminar flows, both steady and unsteady.
\cite{J22} derived the
equations of motion for an inertialess ellipsoid
in a steady uniform shear flow at low Reynolds number.
For a spheroid (i.e. an ellipsoid of revolution), Jeffery showed that
the axis of symmetry of the particle performs a periodic motion on a closed
orbit.
\citet{B62} subsequently extended
Jeffery's analysis to particles of a more general shape: he demonstrated 
that, except
for certain very long particles, the dynamics of any body of revolution 
transported by a low-Reynolds-number shear is equivalent to that of a 
spheroid with an effective aspect ratio.
Amongst bodies of revolution, 
rigid dumbbells and rods have received a systematic investigation
\citep*{BWE71,DE86}.

In Jeffery's \citeyearpar{J22} and Bretherton's \citeyearpar{B62} derivations, 
not only fluid and particle inertia are disregarded, but also Brownian
fluctuations due to the collisions of the particle
with the molecules of the fluid. 
Nevertheless, if
a particle is sufficiently small,
molecular diffusion does influence its orientational dynamics, 
as was shown by \citet{LH71} and \citet{HL72}. 
The review article by
\cite{B74} collects
analytical results on the motion of rigid neutrally buoyant
bodies of revolution subject to a uniform velocity gradient
and to Brownian fluctuations.
A more recent review on this problem can be found in \citet{P99}.
Finally, even in simple laminar flows, the orientation of
particles can form complex patterns; this behaviour was studied
by \citet{S93} in the context of the theory of dynamical systems.

In turbulent flows and in some chaotic flows, the velocity gradient
exhibits small-scale fluctuations. Thus, depending on the size of the
particles, Jeffery's assumption of a spatially uniform velocity gradient
may not be applicable.
To measure the probability of orientations,
experiments have generally used relatively large particles, and therefore 
the velocity gradient could vary appreciably over the size of a particle
\citep*[e.g.][]{KG88,BS94,NB98,PBA05,PGKOGV11,ZGBVPP11a,ZGBVPP11b}.
Accordingly, theoretical studies were mainly concerned with the derivation 
of model equations for the probability of orientations, in which
turbulent fluctuations were
treated as an effective isotropic diffusion term \citep{OK98,SK05}.
Jeffery's approach, however, remains 
applicable to chaotic or turbulent
flows provided that
the particles are sufficiently small. 
The orientation dynamics of tiny elongated particles was
studied numerically in channel flows \citep*{ZAFM01,MAGB08a, MAGB08b}, 
in isotropic turbulence \citep{SK05,PW11,PCTV12}, 
and in chaotic velocity fields \citep*{WBM09}.
In contrast with the case of laminar flows,
few analytical results seem to exist
for the probability distribution of orientations of small particles
transported by a turbulent or chaotic flow.
\citet{T07} examined
the tumbling motion of rodlike polymers in a random flow 
resulting from the superposition of a mean shear and of
white-in-time isotropic fluctuations. 
\citet{WK12} recently studied the alignment of rods with vorticity
in a turbulent isotropic flow.

Here, the probability density function (p.d.f.)
of orientations is derived exactly
for small particles transported by a random flow with
axisymmetric statistics. The particles are general bodies of revolution
possessing fore-aft symmetry.
The axisymmetry of the flow means 
that the velocity field is statistically invariant
under arbitrary rotations of the axes about a given direction
as well as under reflections in planes containing that direction
or normal to it \citep{B46,C50}.
Axisymmetry is the simplest form of statistical anisotropy \citep{BP05,CBB12}, 
and is found in 
rotating, stratified, or wind-tunnel turbulence \citep{L95}.
Furthermore, the random flow is assumed to be Gaussian and  
to have zero correlation time. The assumption of temporal decorrelation
is adequate when the correlation time of the flow is
short compared to the characteristic time scale of
material-line-element stretching. 
This assumption, albeit restrictive,  
allows a fully analytical solution of the problem.

The evolution equation for the orientation vector of a non-spherical
rigid particle is introduced in~\S~\ref{sec:particles+flow}.
Section~\ref{sec:FPE} is devoted to 
the derivation of the Fokker--Planck equation
for the p.d.f. of the orientation angle. Its stationary solution
is studied in~\S~\ref{sec:stationary}.
Some conclusions are drawn in~\S~\ref{sec:conclusions}.

\section{Orientation dynamics}
\label{sec:particles+flow}

The particles considered here are
rigid bodies of revolution possessing fore-aft symmetry
(although in the literature
such particles are commonly referred to as ``axisymmetric'',
this terminology will be avoided here not to generate confusion; 
the term ``axisymmetry''
will be reserved to the statistical invariance of the velocity field).
The particles are of uniform composition and are suspended in a Newtonian
fluid of the same density. 
Furthermore, the inertia of the particles as well as hydrodynamic
particle-particle interactions are disregarded, and
no externally imposed 
force or couple influences the dynamics.
In particular, it is appropriate to disregard hydrodynamic interactions 
when the suspension is sufficiently dilute.

The undisturbed motion of the fluid is described by 
the velocity field~$\boldsymbol{v}(\boldsymbol{x},t)$.
The size of the particles is assumed to be small compared to the 
typical length over which
the velocity gradient~$\bnabla\boldsymbol{v}=
(\upartial_j v_i)_{1\leqslant i,j\leqslant 3}$ changes
($\upartial_j\equiv\upartial/\upartial x_j$).
For turbulent flows, this assumption is satisfied if
the particles are smaller than
the Kolmogorov scale, where viscosity smooths out the velocity fluctuations;
the Reynolds number of the flow at the scale of a particle
is thus less than~1.
Given their small size, the particles also 
experience Brownian collisions with the molecules of the fluid.

In a sufficiently dilute suspension,
attention can be restricted to the dynamics of a single isolated particle.
The configuration of a body of revolution
is determined by the position of its
centre of mass, $\boldsymbol{r}_c(t)$, and by the orientation of its axis of
revolution, which is specified by a unit vector~$\boldsymbol{N}(t)$ parallel
to the axis itself.
As the particles are subject to Brownian fluctuations,
their dynamics is random even in a laminar flow.
Consider first a deterministic velocity field or a given realisation of
a random velocity field.
On the above assumptions, the centre of mass moves 
according to the following equation \citep[e.g.][]{DE86}:
\begin{equation}
\label{eq:center-of-mass}
\dot{\boldsymbol{r}}_c(t)
=\boldsymbol{v}(\boldsymbol{r_c}(t),t)+\sqrt{\mathcal{D}_T}
\,\boldsymbol{\zeta}(t), 
\end{equation}
where~$\mathcal{D}_T>0$ is the translational diffusion coefficient
and~$\boldsymbol{\zeta}(t)$ 
is three-dimensional white noise, i.e.
a Gaussian stochastic process with
\begin{equation}
\langle\boldsymbol{\zeta}(t)\rangle=0 \qquad \mbox{and}\quad
\langle\zeta_i(t+\tau)\zeta_j(t)\rangle=\delta_{ij}\delta(\tau)
\end{equation}
for all~$t,\tau>0$ and $i,j=1,2,3$. 
The orientation vector satisfies the following stochastic
differential equation (summation over repeated indexes is implied):
\begin{equation}
\label{eq:N}
\skew3\dot{N}_i=\kappa_{ij}(t)N_j-
\kappa_{pq}(t)\dfrac{N_pN_q}{\vert\boldsymbol{N}\vert^2}N_i
+\sqrt{\mathcal{D}_R}\,\varSigma_{ij}(\boldsymbol{N})\circ\xi_j(t),
\qquad \vert\boldsymbol{N}(0)\vert=1,
\end{equation}
where
\begin{equation}
\label{eq:kappa}
\boldsymbol{\kappa}(t)=\mathsfbi{\Omega}(t)+\gamma\mathsfbi{E}(t)
\end{equation}
with
\begin{equation}
\label{eq:Omega+E}
\mathsfbi{\Omega}(t)=
\dfrac{\mathsfbi{G}(t)-\mathsfbi{G}^{\mathrm{T}}(t)}{2},
\qquad
\mathsfbi{E}(t)=\dfrac{\mathsfbi{G}(t)+\mathsfbi{G}^{\mathrm{T}}(t)}{2},
\end{equation}
and~$\mathsfbi{G}(t)=\bnabla\boldsymbol{v}(\boldsymbol{r}_c(t),t)$.
Thus, $\mathsfbi{\Omega}(t)$ and~$\mathsfbi{E}(t)$
are the vorticity tensor
and the rate-of-strain tensor evaluated at~$\boldsymbol{r}_c(t)$
and $\boldsymbol{\kappa}(t)$ is an effective Lagrangian velocity gradient.
The scalar constant~$\gamma$ depends on the geometrical shape of the particle.
For~$\vert \gamma\vert<1$, the evolution equation can be mapped into that 
of a spheroid with aspect ratio equal to~$\sqrt{(1+\gamma)/(1-\gamma)}$
\citep{B62}. 
Prolate spheroids are obtained for $0<\gamma<1$, 
oblate spheroids for $-1<\gamma<0$. 
Special cases are: 
spheres ($\gamma=0$),
rigid dumbbells ($\gamma=1$),
rods ($\gamma=1$), and
disks ($\gamma=-1$). Furthermore,
it was shown by~\citet{B62} that in principle there exist very long particles
for which~$\vert\gamma\vert>1$.
In~\eqref{eq:N}, the random vector~$\boldsymbol{\xi}(t)$ 
is three-dimensional white noise and hence has the same properties
as~$\boldsymbol{\zeta}(t)$, but is statistically independent of it. 
The matrix~$\mathsfbi{\Sigma}(\boldsymbol{n})$ has the following form:
\begin{equation}
\label{eq:sigma}
\mathsfbi{\Sigma}(\boldsymbol{n})
={\mathsfbi{I}}-\boldsymbol{n}\boldsymbol{n}/\vert\boldsymbol{n}\vert^2
\end{equation}
and~$\mathcal{D}_R>0$ is the rotary diffusion coefficient.
By using the Cauchy--Schwarz inequality, it is easy to check 
that~$ \mathsfbi{\Sigma}(\boldsymbol{n})$ is positive semi-definite.
The symbol~$\circ$ indicates that the stochastic term in~\eqref{eq:N}
is understood in the Stratonovich sense.

Equation~\eqref{eq:N} is Jeffery's equation for the orientation vector
of a body of revolution with the addition of a stochastic term modelling
Brownian fluctuations. 
The stochastic term is chosen in
such a way as to produce isotropic diffusion of~$\boldsymbol{N}(t)$ 
on the unit sphere, so that~$\vert \bm N(t)\vert$ is preserved in time
(see appendix~\ref{app:A} for more details).
It is worth remarking that a Brownian term of the same form
has been used to
model the turbulent fluctuations of the velocity gradient 
\citep{KG88,OK98,SK05}
or to describe particle--particle interactions both in semi-dilute suspensions
\citep{DE86} and in concentrated suspensions \citep{DE78,KD80}.
 
Equation~\eqref{eq:N} can be generalised to the case of a homogeneous
axisymmetric random flow.
The velocity field transporting the particle
is Gaussian and has zero mean and correlation:
\begin{equation}
\label{eq:correlation-v}
\langle v_{i}(\boldsymbol{x}+\boldsymbol{r},t+\tau)
v_j(\boldsymbol{x}, t)\rangle
=Q_{ij}(\boldsymbol{r})\delta(\tau), \qquad i,j=1,2,3.
\end{equation}
The form of the
correlation guarantees that~$\boldsymbol{v}(\boldsymbol{x},t)$ is statistically
homogeneous in space. Additionally, the velocity field is assumed to be
incompressible 
($\bnabla\bcdot\boldsymbol{v}=0$) and statistically axisymmetric
with respect to the
direction specified by the unit vector~$\boldsymbol{\lambda}$.
The tensor~$\mathsfbi{Q}(\boldsymbol{r})$ must then take the 
form:
\begin{equation}
\label{eq:q}
Q_{ij}(\bm r)=A r_i r_j+
B\delta_{ij}+C\lambda_i\lambda_j
+D(r_i\lambda_j+\lambda_i r_j),
\end{equation}
where~$A$, $B$, $C$, $D$ are smooth functions
of~$\vert\boldsymbol{r}\vert^2$ 
and~$(\boldsymbol{r}\bcdot\boldsymbol{\lambda})$; $A$, $B$, and~$C$ are even
in~$(\boldsymbol{r}\bcdot\boldsymbol{\lambda})$,
while~$D$ is odd in~$(\boldsymbol{r}\bcdot\boldsymbol{\lambda})$ 
\citep{B46,C50}.
Furthermore, the functions~$A$, $B$, $C$, $D$ are not independent and
satisfy certain differential relations \citep{B46,C50}.
The velocity field defined above
is an axisymmetric generalisation
of the isotropic random flow
introduced by \cite{K68} in the context of passive turbulent
transport. The same axisymmetric
velocity field was used by \citet{SK92}
to study polymer stretching in flows through random beds of fibres.

The velocity gradient is also Gaussian and zero-mean;
the single-point two-time
correlation can be derived by
using the statistical homogeneity of the velocity field:
\begin{equation}
\label{eq:correlation-gradient}
\langle\upartial_j v_i(\bm x,t+\tau)\upartial_q v_p(\bm x, t)\rangle
=\varGamma_{ijpq}\delta(\tau)
\qquad
i,j,p,q=1,2,3
\end{equation}
with
\begin{equation}
\label{eq:corr-grad}
\varGamma_{ijpq}=
-\left.\frac{\upartial^2 Q_{ip}}{\upartial r_j\upartial r_q}
\right\vert_{\boldsymbol{r}=0}.
\end{equation}
Substituting~\eqref{eq:q} in~\eqref{eq:corr-grad} yields
(see equation~(5.12) in \citet{B46}):
\begin{equation}
\begin{split}
\varGamma_{ijpq}
=&\,(d+4a)\delta_{jq}\delta_{ip}-a(\delta_{pq}\delta_{ij}
+\delta_{iq}\delta_{pj})
+(b+c+5d)\delta_{jq}\lambda_i\lambda_p
-b\delta_{ip}\lambda_j\lambda_q
\\
&-d[(\delta_{pq}\lambda_i+\delta_{iq}\lambda_p)\lambda_j
+(\delta_{pj}\lambda_i+\delta_{ij}\lambda_p)\lambda_q]
-c\lambda_i\lambda_p\lambda_j\lambda_q,
\end{split}
\end{equation}
where~$a$, $b$, $c$, $d$ are real constants.
For the sake of simplicity,
$\boldsymbol{\lambda}$ is taken
in the direction of the third axis, i.e. $\boldsymbol{\lambda}
=(0,0,1)$. Then, the coefficients in~$\varGamma_{ijpq}$ are written:
\begin{equation}
\begin{array}{rcrrcl}
2a&=&\varGamma_{1212}-\varGamma_{1111}&
\qquad
d&=&\varGamma_{3333}-\varGamma_{1111}
\\[3mm]
b&=&\varGamma_{1212}-\varGamma_{1313}&
\qquad 2a+b+c+4d&=&\varGamma_{3131}-\varGamma_{3333}.
\end{array}
\end{equation}
The two-time correlation of the components of the vorticity~$\bm\omega=
\bnabla\times\bm v$ can be expressed in terms of~$a$, $c$, and~$d$
as follows \citep[][p.~490]{B46}:
\begin{equation}
\label{eq:vorticity-1}
\langle\omega_1(\bm x,t+\tau)\omega_1(\bm x,t)\rangle=
\langle\omega_2(\bm x,t+\tau)\omega_2(\bm x,t)\rangle
=(10a+c+9d)\delta(\tau)
\end{equation}
and
\begin{equation}
\langle\omega_3(\bm x,t+\tau)\omega_3(\bm x,t)\rangle
=(10a+2d)\delta(\tau).
\end{equation}
The coefficients~$a$, $b$, $c$, $d$ are not free; they are
constrained by
the following inequalities
(see appendix~\ref{app:B}):
\begin{subeqnarray}
\label{eq:inequalities}
\slabel{eq:ineq-1}
&\begin{cases}
a+d> 0 & \text{if $a\ge 0$}, \\
5a+d> 0 & \text{if $a<0$},
\end{cases}&
\\
\slabel{eq:ineq-2}
&4a-b+d> 0, \qquad 4a+b+c+6d> 0,&
\\
\slabel{eq:ineq-3}
&15a^2-b^2-(b-d)(c+5d)+2a(2c+13d)> 0.&
\end{subeqnarray}
If~$a>0$ and~$b=c=d=0$, then~$\varGamma_{ijpq}$
gives the single-point correlation of the gradient of an 
isotropic velocity field~\citep{R40}.
Thus, $a$ determines the intensity of the isotropic part
of the gradient and~$b$, $c$, $d$ control the
statistical anisotropy of the flow. 

The orientation dynamics of the particle depends on the velocity gradient
evaluated at~$\boldsymbol{r}_c(t)$, which was denoted as~$\mathsfbi{G}(t)$
in \S~\ref{sec:particles+flow}.
In virtue of the $\delta$-correlation in time and the statistical
homogeneity of the flow, $\mathsfbi{G}(t)$
has the same temporal statistics as~$\bnabla\boldsymbol{v}
(\boldsymbol{x},t)$ for any
given~$\boldsymbol{x}$~\citep*{FGV01}
(the presence of white noise in~\eqref{eq:center-of-mass}
does not modify the statistics of~$\mathsfbi{G}(t)$).
The components of~$\boldsymbol{\kappa}(t)$, defined in~\eqref{eq:kappa},
are a linear combination of the components of~$\mathsfbi{G}(t)$, and
consequently~$\boldsymbol{\kappa}(t)$ is a Gaussian process with zero mean
and correlation:
\begin{equation}
\label{eq:correlation-K}
\langle \kappa_{ij}(t+\tau)\kappa_{pq}(t)\rangle=
K_{ijpq}\delta(\tau)
\end{equation}
with
\begin{equation}
K_{ijpq}=
\dfrac{1}{4}[\varGamma_{ijpq}-\varGamma_{ijqp}
-\Gamma_{jipq}+\varGamma_{jiqp}+2\gamma
(\varGamma_{ijpq}-\varGamma_{jiqp})+\gamma^2
(\varGamma_{ijpq}+\varGamma_{ijqp}
+\varGamma_{jipq}+\varGamma_{jiqp})].
\end{equation}
The form of~$K_{ijpq}$ can be derived by 
substituting~\eqref{eq:kappa} and~\eqref{eq:Omega+E} into
the left-hand side of~\eqref{eq:correlation-K}
and by using~\eqref{eq:correlation-gradient}.
The tensor~$K_{ijpq}$ is positive semi-definite, for
it is the covariance of a second-order Gaussian tensor.
Thus,
in the evolution equation for the orientation vector,
$\boldsymbol{\kappa}(t)$ plays the role
of a multiplicative tensorial white noise.
As~$\boldsymbol{\kappa}(t)$ can be thought of as an approximation
of a real noise process in the limit of zero correlation time,
the corresponding terms in~\eqref{eq:N} must be interpreted
in the Stratonovich sense \citep[e.g.][p.~227]{KP92}. 
The stochastic differential
equation for the orientation vector of a non-spherical particle can then
be rewritten as follows:
\begin{equation}
\label{eq:components}
\skew3\dot{N}_i=M_{ipq}(\boldsymbol{N})\circ\kappa_{pq}(t)
+\sqrt{\mathcal{D}_R}\,\varSigma_{ij}(\boldsymbol{N})\circ\xi_j(t),
\end{equation}
where~$M_{ipq}(\boldsymbol{n})=
(\delta_{ip}\delta_{jk}-\delta_{ik}\delta_{jp})n_jn_kn_q/
\vert\boldsymbol{n}\vert^2$
and the initial condition~$\boldsymbol{N}(0)$ 
is such that~$\vert\boldsymbol{N}(0)\vert=1$.
%
%

\section{Fokker--Planck equation for the probability density
function of the orientation angle}
\label{sec:FPE}
As~$\boldsymbol{v}(\boldsymbol{x},t)$ is statistically invariant
under spatial translations, 
the p.d.f.
of~$\boldsymbol{N}(t)$ taking the value~$\boldsymbol{n}=(n_1,n_2,n_3)$ at
time~$t$ is independent of~$\boldsymbol{r}_c$ and
is thus denoted by~$f(\boldsymbol{n};t)$.
The It\^o equation equivalent to~\eqref{eq:components} is
\begin{equation}
\label{eq:components-Ito}
\skew3\dot{N}_i=\beta_i(\boldsymbol{N})+M_{ipq}(\boldsymbol{N})
\kappa_{pq}(t)
+\sqrt{\mathcal{D}_R}\,\varSigma_{ij}(\boldsymbol{N})\xi_j(t)
\end{equation}
with
\begin{equation}
\beta_{i}(\boldsymbol{n})=
\dfrac{1}{2}\,K_{mnpq}
M_{jpq}(\boldsymbol{n})\dfrac{\upartial}{\upartial n_j}
M_{imn}(\boldsymbol{n})
+\dfrac{\mathcal{D}_R}{2}\varSigma_{jk}(\boldsymbol{n})
\dfrac{\upartial}{\upartial n_j}\varSigma_{ik}(\boldsymbol{n}).
\end{equation}
Consequently,
$f(\boldsymbol{n};t)$ satisfies the Fokker--Planck equation:
\begin{equation}
\label{eq:FPE-cartesian}
\dfrac{\upartial f}{\upartial t}=-
\dfrac{\upartial}{\upartial n_i}\, [\beta_i(\bm n)f]
+\dfrac{1}{2}\,\dfrac{\upartial^2}{\upartial n_i
\upartial n_j}\,[\alpha_{ij}(\bm n)f]
%
\end{equation}
with
\begin{equation}
\begin{split}
\alpha_{ij}(\bm n)
&=K_{mnpq}M_{imn}(\bm n)M_{jpq}(\bm n)
+\mathcal{D}_R\varSigma_{ik}(\bm n)\varSigma_{jk}(\bm n)
\\
&=K_{mnpq}M_{imn}(\bm n)M_{jpq}(\bm n)
+\mathcal{D}_R\varSigma_{ij}(\bm n),
\end{split}
\end{equation}
where the last equality follows from \eqref{eq:properties-sigma}.
Equations \eqref{eq:components-Ito} and \eqref{eq:FPE-cartesian} 
can be derived from~\eqref{eq:components}
by using the formal rules~$\kappa_{ij}(t)dt=O(\sqrt{dt})$
and~$\kappa_{ij}(t)dt\,\kappa_{pq}(t)dt=K_{ijpq}dt$ and by proceeding as
in the case of a vectorial white noise (see \citealt{G83} and the appendix in
\citealt{FGV01}). The diffusion tensor~$\boldsymbol{\alpha}$ is positive
semi-definite as a consequence of the positive semi-definiteness
of~$\mathsfbi{K}$ and~$\mathsfbi{\Sigma}$.

\begin{figure}
\centering
\includegraphics[width=7cm]{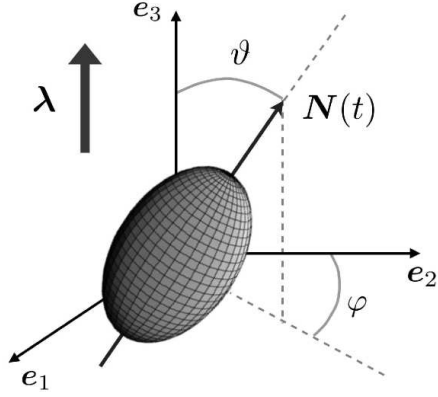}
\caption{Orientation of a non-spherical particle.}
\label{fig:dumbbell}
\end{figure}

To study the orientation dynamics of a non-spherical particle, 
it is convenient to move from Cartesian coordinates~$(n_1,n_2,n_3)$
to spherical coordinates~$(n,\vartheta,\varphi)$ according to the
usual transformations:
\begin{equation}
\label{eq:spherical}
n=\sqrt{n_1^2+n_2^2+n_3^2},\qquad\vartheta=
\arctan\Big(\sqrt{n_1^2+n_2^2}/n_3\Big),
\qquad\varphi=\arctan(n_2/n_1)
\end{equation}
with~$0\leq n$, $0\le\vartheta\le \pi$, and $0\le\varphi<2\pi$ 
(figure~\ref{fig:dumbbell}).
On account of the fixed length of the orientation vector,
the probability den\-si\-ty func\-tion of orientations must
take the form $f(n,\vartheta,\varphi;t)=
\psi(\vartheta,\varphi;t)\delta(n-L)$ with~$L=1$. 
Thus, $\psi(\vartheta,\varphi;t)\sin\vartheta\,
\mathrm{d}\vartheta\,\mathrm{d}\varphi$ 
is the probability of the particle being oriented at time~$t$ within
an elementary solid angle~$\sin\vartheta\,\mathrm{d}\vartheta\,\mathrm{d}\varphi$
of~$(\vartheta,\varphi)$. In addition, the following normalisation holds:
\begin{equation}
\int_0^{\upi} \int_0^{2\upi}
\psi(\vartheta,\varphi;t)\sin\vartheta\,
\mathrm{d}\vartheta\,\mathrm{d}\varphi=1
\qquad \forall\; t\geqslant 0.
\end{equation}
The function~$\varPsi(\vartheta,\varphi;t)=
\psi(\vartheta,\varphi;t)\sin\vartheta$
satisfies a new Fokker--Planck equation, which
can be derived
from~\eqref{eq:FPE-cartesian} by using the transformation formulae
for the drift and diffusion coefficients under a change of variables
\citep[][p.~88]{R89}, by writing~$f(n,\vartheta,\varphi;t)=
\psi(\vartheta,\varphi;t)\delta(n-L)$ with~$L=1$, and 
by integrating the resulting equation with respect to~$n$.
The final result is:
\begin{equation}
\label{eq:FPE-spherical}
\dfrac{\upartial \varPsi}{\upartial t}= 
-\frac{\upartial}{\upartial\vartheta}
[\mathcal{B}_{\vartheta}(\vartheta)\varPsi]+
\frac{1}{2}\frac{\upartial^2}{\upartial\vartheta^2}
[\mathcal{A}_{\vartheta\vartheta}(\vartheta)\varPsi]+
\frac{1}{2}\mathcal{A}_{\varphi\varphi}(\vartheta)
\frac{\upartial^2 \varPsi}{\upartial\varphi^2},
\end{equation}
where
\begin{subeqnarray}
\label{eq:b_theta}
\mathcal{B}_\vartheta(\vartheta)&=& \beta_{i}(\bm n)\,
\dfrac{\partial \vartheta}{\partial n_i}
+\dfrac{\alpha_{ij}(\bm n)}{2}\,
\dfrac{\partial^2\vartheta}{\partial n_i\partial n_j}
=\left[\mu_1+\mu_2\sin^2(\vartheta)+
\mu_3\sin^4(\vartheta)\right]\cot\vartheta,
\\[1mm]
\label{eq:a_theta_theta}
\mathcal{A}_{\vartheta\vartheta}(\vartheta)&=&
\alpha_{ij}(\bm n)\,
\dfrac{\partial\vartheta}{\partial n_i}\,
\dfrac{\partial\vartheta}{\partial n_j}
=2\mu_1+\mu_4\sin^2(\vartheta)+\mu_3\sin^4(\vartheta),
\\[1mm]
\label{eq:a_phi_phi}
\mathcal{A}_{\varphi\varphi}(\vartheta)&=&
\alpha_{ij}(\bm n)\, 
\dfrac{\partial\varphi}{\partial n_i}\,
\dfrac{\partial\varphi}{\partial n_j}
=\mu_5+2\mu_1\csc^2(\vartheta)
\end{subeqnarray}
with
\begin{subeqnarray}
\mu_1 &=& \textstyle\frac{1}{8}\left[2(5+3\gamma^2)a-4\gamma b+(1-\gamma)^2 c
+(9-10\gamma+5\gamma^2)d+4\mathcal{D}'_R\right],
\\
\mu_2 &=&\textstyle\frac{3}{4}\gamma\left[2 b+(1-\gamma) c+
(5-\gamma) d\right],
\\
\mu_3&=& c\gamma^2,
\\
\mu_4&=&\gamma[2b+(1-\gamma)c+5d],
\\
\mu_5 &=& \textstyle\frac{1}{4}
\left[4b\gamma-(1-\gamma)^2c-(1-\gamma)(7-3\gamma)d\right],
\end{subeqnarray}
and~$\mathcal{D}'_R=\mathcal{D}_R/L^2$ with~$L=1$ (the numerical values 
of~$\mathcal{D}'_R$ and~$\mathcal{D}_R$ coincide, but their physical
dimensions are different).
Equations~\eqref{eq:b_theta}
involve the Jacobian and the Hessian of the transformation from Cartesian to
spherical coordinates, which can be calculated from~\eqref{eq:spherical}.
Inequalities~\eqref{eq:inequalities} guarantee
that~$\mathcal{A}_{\vartheta\vartheta}(\vartheta)$
and~$\mathcal{A}_{\varphi\varphi}(\vartheta)$ are strictly
positive (appendix~\ref{app:B}).

The contribution to~\eqref{eq:FPE-spherical} due to the isotropic part
of~$\mathsfbi{\Gamma}$ is a Laplace--Beltrami term with diffusion coefficient
proportional to~$a$; this contribution is of the same form as that
coming from~$\boldsymbol{\xi}(t)$. The isotropic component of the flow and the
Brownian fluctuations 
therefore have the same effect on the orientation statistics of the particle.

Since~$\vartheta$ and~$\varphi$ are angular variables,
the boundary conditions for~$\varPsi(\vartheta,\varphi;t)$ are periodic:
\begin{equation}
\label{eq:periodic-bc}
\varPsi(\vartheta,\varphi;t)=\varPsi(\vartheta+2\upi,\varphi;t)
\qquad \text{and}\qquad
\varPsi(\vartheta,\varphi;t)=\varPsi(\vartheta,\varphi+2\upi;t)
\end{equation}
for all~$\vartheta$, $\varphi$, and~$t$.
The long-time properties of~$\varPsi(\vartheta,\varphi;t)$ 
can be deduced from~\eqref{eq:FPE-spherical}. If the partial derivative
with respect to time is dropped, then 
\eqref{eq:FPE-spherical} is invariant under
the transformations~$\vartheta\leftrightarrow 2\upi-\vartheta$
(reflections with respect to planes containing~$\boldsymbol{\lambda}$)
and~$\vartheta\leftrightarrow\upi-\vartheta$ (reflections with respect to
planes orthogonal to~$\boldsymbol{\lambda}$). Moreover,
the coefficients~$\mathcal{B}_\vartheta$, $\mathcal{A}_{\vartheta\vartheta}$,
and~$\mathcal{A}_{\varphi\varphi}$ do not depend on~$\varphi$
(invariance under rotations about~$\boldsymbol{\lambda}$).
These properties of~\eqref{eq:FPE-spherical}
are a natural consequence of the statistical axisymmetry
of the velocity field and translate into analogous properties of the
stationary p.d.f. of orientations.

The invariance of~\eqref{eq:FPE-spherical}
under rotations about~$\boldsymbol{\lambda}$ can be used to derive
a one-dimensional Fokker--Planck equation for
the marginal p.d.f.: $\widehat{\varPsi}(\vartheta;t)
=\widehat{\psi}(\vartheta;t)\sin\vartheta$
with $\widehat{\psi}(\vartheta;t)=\int_0^{2\upi}
\psi(\vartheta,\varphi;t)\mathrm{d}\varphi$. 
Integrating~\eqref{eq:FPE-spherical} with respect to~$\varphi$ 
from~$0$ to~$2\upi$ and making use of~\eqref{eq:periodic-bc} yield:
\begin{equation}
\label{eq:FPE-theta}
\dfrac{\upartial\widehat{\varPsi}}{\upartial t}= 
-\frac{\upartial}{\upartial\vartheta}
\left[\mathcal{B}_{\vartheta}(\vartheta)\widehat{\varPsi}\right]+
\frac{1}{2}\frac{\upartial^2}{\upartial\vartheta^2}
\left[\mathcal{A}_{\vartheta\vartheta}(\vartheta)\widehat{\varPsi}\right].
\end{equation}
The solution of the above equation
must be normalised and periodic:
$\widehat{\varPsi}(\vartheta;t)=\widehat{\varPsi}(\vartheta+2\upi;t)$
for all~$\vartheta,t$. 
A direct consequence of~\eqref{eq:FPE-theta} is
that, along the trajectory of the particle,
the time evolution of~$\vartheta(t)$ is decoupled
from that of~$\varphi(t)$ and is described by the stochastic ordinary
differential equation:
\begin{equation}
\label{eq:theta}
\skew3\dot{\vartheta}(t)=\mathcal{B}_\vartheta(\vartheta(t))+
\sqrt{\mathcal{A}_{\vartheta\vartheta}(\vartheta(t))}\,\eta(t),
\qquad 0\leqslant\vartheta(t)\leqslant\upi,
\end{equation}
where~$\eta(t)$ is white noise.

\section{Stationary statistics of the orientation angle}
\label{sec:stationary}

It was argued in \S~\ref{sec:FPE} that
the stationary solution of~\eqref{eq:FPE-spherical}, $\varPsi_{\mathrm{st}}$, 
does not depend on~$\varphi$. Therefore, $\varPsi_{\mathrm{st}}$
solves the equation:
\begin{equation}
\dfrac{1}{2}\,\dfrac{d^2}{d\vartheta^2}
\left[\mathcal{A}_{\vartheta\vartheta}(\vartheta)\varPsi_{\mathrm{st}}\right]
-\dfrac{d}{d\vartheta}
\left[\mathcal{B}_{\vartheta}(\vartheta)\varPsi_{\mathrm{st}}\right]=0.
\end{equation}
Thanks to the periodic boundary conditions~\eqref{eq:periodic-bc},
$\varPsi_{\mathrm{st}}(\vartheta)$ takes the following simple form
(see appendix~\ref{app:C}):
\begin{equation}
\label{eq:Psi-stationary}
{\varPsi}_{\mathrm{st}}(\vartheta)=
\dfrac{\mathcal{N}}{\mathcal{A}_{\vartheta\vartheta}(\vartheta)}\,
\exp\left[2\int_{\vartheta_0}^\vartheta\dfrac{\mathcal{B}_\vartheta(z)}%
{\mathcal{A}_{\vartheta\vartheta}(z)}\,\mathrm{d}z\right],
\end{equation}
where~$0\leqslant\vartheta_0\leqslant\upi$ 
and~$\mathcal{N}$ is a normalisation constant such that
\begin{equation}
\label{eq:normalisation}
2\upi\int_0^{\upi} \varPsi_{\mathrm{st}}(\vartheta)\mathrm{d}\vartheta=1.
\end{equation}
In~\eqref{eq:Psi-stationary}, the choice of~$\vartheta_0$ is
in fact unimportant, since it only modifies the normalisation constant.

Only the case~$\mu_3\neq 0$ is considered here; the case~$\mu_3=0$
is examined in appendix~\ref{app:D}, even though no new physical regimes 
emerge in this latter case.
For~$\mu_3\neq 0$, the integral in~\eqref{eq:Psi-stationary}
can be calculated by using the change of variable~$y=\sin^2(\omega)$ and
formulae~2.172, 2.175(1), and~2.177(1) of \cite{GR65}.
The final result is:
\begin{equation}
\label{eq:psi-stationary}
{\psi}_{\mathrm{st}}(\vartheta)=\frac{\mathcal{N} \chi(\vartheta)}%
{[\mathcal{A}_{\vartheta\vartheta}(\vartheta)]^{3/4}},
\end{equation}
where
\begin{equation}
\label{eq:chi}
\chi(\vartheta)=\begin{cases}
\exp\left\{\dfrac{2}{\sqrt{\varDelta}}\left(\mu_2-\dfrac{3}{4}\mu_4\right)
\arctan\left[\dfrac{\mu_4+2\mu_3\sin^2(\vartheta)}{\sqrt{\varDelta}}\right]
\right\} & (\varDelta>0)
\\[5mm]
\exp\left\{\left(\dfrac{3}{4}\mu_4-\mu_2\right)
\dfrac{2}{\mu_4+2\mu_3\sin^2(\vartheta)}\right\} & (\varDelta=0)
\\[5mm]
\left\vert\dfrac{\sqrt{-\varDelta}+\mu_4+2\mu_3\sin^2(\vartheta)}%
{\sqrt{-\varDelta}-\mu_4-2\mu_3\sin^2(\vartheta)}\right\vert^{%
\frac{1}{\sqrt{-\varDelta}}\left(\frac{3}{4}\mu_4-\mu_2\right)}
& (\varDelta<0)
\end{cases}
\end{equation}
with~$\varDelta=8\mu_1\mu_3-\mu_4^2$.
It is shown in appendix~\ref{app:B} that~$\psi_{\mathrm{st}}(\vartheta)$
is bounded for all values of~$a$, $b$, $c$, $d$, 
and~$\gamma$. The stationary p.d.f. of orientations
satisfies~${\psi}_{\mathrm{st}}(2\upi-\vartheta)=
{\psi}_{\mathrm{st}}(\vartheta)$ 
and~${\psi}_{\mathrm{st}}(\upi-\vartheta)=
{\psi}_{\mathrm{st}}(\vartheta)$. These properties
are a consequence of the statistical symmetries of the carrier flow,
as was noted after~\eqref{eq:periodic-bc}.

Since~$\varDelta$ can be written 
as~$\varDelta=a^2F(\gamma,b/a,c/a,d/a,\mathcal{D}'_R/a)$, 
${\psi}_{\mathrm{st}}(\vartheta)$ only depends
on the ratios~$b/a$, $c/a$, $d/a$, $\mathcal{D}'_R/a$ 
(and on~$\gamma$). The same conclusion
could have been reached by rescaling~$t$ by~$a^{-1}$ 
in~\eqref{eq:FPE-spherical}.

For a spherical particle (i.e. $\gamma=0$), $\mu_2$ and~$\mu_4$ vanish.
Consequently,~$\chi(\vartheta)=1$, 
$\mathcal{A}_{\vartheta\vartheta}(\vartheta)=\mathrm{const.}$, 
and hence~${\psi}_{\mathrm{st}}(\vartheta)=(4\upi)^{-1}$, 
in accordance with the fact that all orientations are equally probable for
a sphere.
Similarly, if~$a$ or~$\mathcal{D}'_R$ are much greater than~$b$,
$c$, $d$, i.e. if the isotropic component of the velocity field
or the Brownian fluctuations prevail on the anisotropic component
of the flow, then~$\mu_1\gg\mu_i$, $i=2,\dots,5$ and
${\psi}_{\mathrm{st}}(\vartheta)$ weakly depends
on~$\vartheta$ regardless of the shape of the particle.
In the following, therefore, $\gamma$ is assumed to be nonzero 
and~$b$, $c$, $d$ are of the same order 
of magnitude as~$a$ and~$\mathcal{D}'_R$ or greater.

The behaviour of~${\psi}_{\mathrm{st}}(\vartheta)$
can be deduced from that of its first derivative.
By using~$\mathcal{A}_{\vartheta\vartheta}(\vartheta)> 0$,
it can be shown that, for all values of~$\varDelta$,
\begin{equation}
\label{eq:derivative}
\dfrac{\mathrm{d}}{\mathrm{d}\vartheta}
{\psi}_{\mathrm{st}}(\vartheta)=\gamma\sin(2\vartheta)
[c\gamma\cos(2\vartheta)-\sigma]h(\vartheta),
\end{equation}
where
\begin{equation}
\sigma\equiv 2b+c+(5+\gamma)d
\end{equation}
and~$h(\vartheta)$ is a strictly positive function
for all~$0\leqslant\vartheta\leqslant\upi$.
Four regimes are then identified depending on the properties of the
axisymmetric flow and on the geometrical shape of the particle:
\begin{enumerate}
\renewcommand{\theenumi}{\roman{enumi}}
\item\emph{Rotation about the axis of symmetry of the flow}.
For the following values of the parameters:
\begin{equation}
\label{eq:rotation}
\vert\sigma\vert>\vert c\gamma\vert
\qquad \text{and} \qquad 
\gamma\sigma<0,
\end{equation}
the function~${\psi}_{\mathrm{st}}$ only has
three extrema in~$\vartheta=0,\upi/2,\upi$ (indeed the equation
$c\gamma\cos(2\vartheta)=\sigma$ has no solution). More precisely,
${\psi}_{\mathrm{st}}(\vartheta)$
has a maximum in~$\upi/2$ and two minima in~$0$ and~$\upi$
(figure~\ref{fig:rotation+tumbling}).
Thus, in this regime
the particle rotates about the direction~$\boldsymbol{\lambda}$; the level of
alignment with the plane perpendicular to~$\boldsymbol{\lambda}$ decreases
as the degree of anisotropy of the flow vanishes or
the shape of the particle approaches the spherical one
(figure~\ref{fig:rotation+tumbling}).
\begin{figure}
\setlength{\unitlength}{\textwidth}
\includegraphics[width=0.495\textwidth]{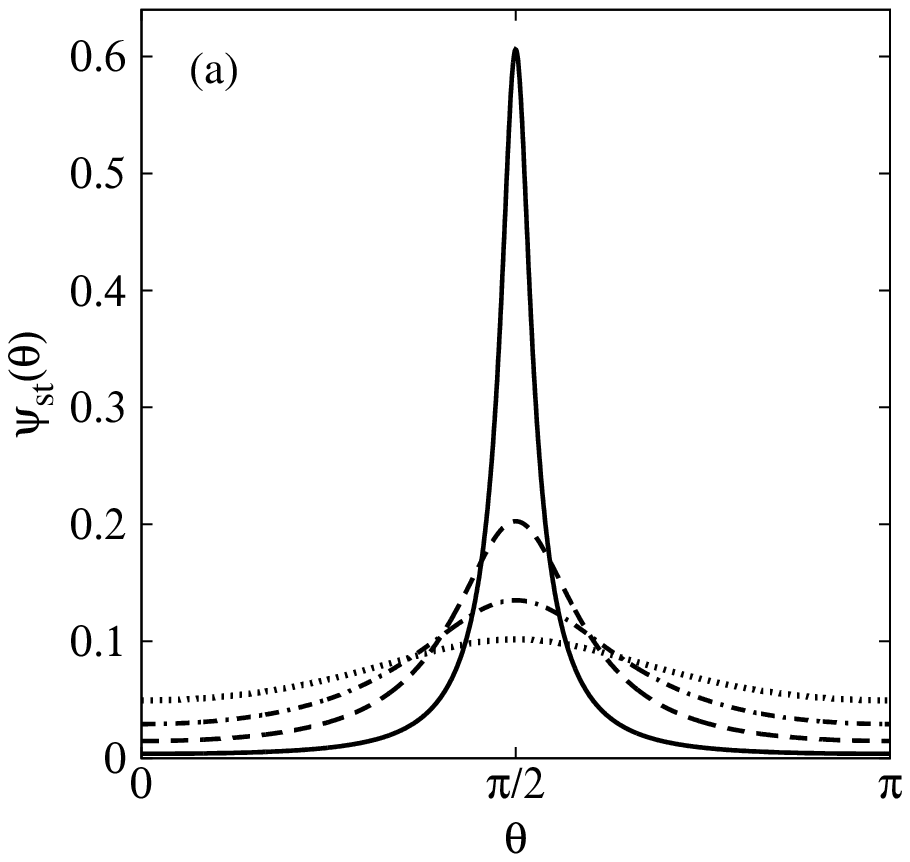}
\hfill
\includegraphics[width=0.49\textwidth]{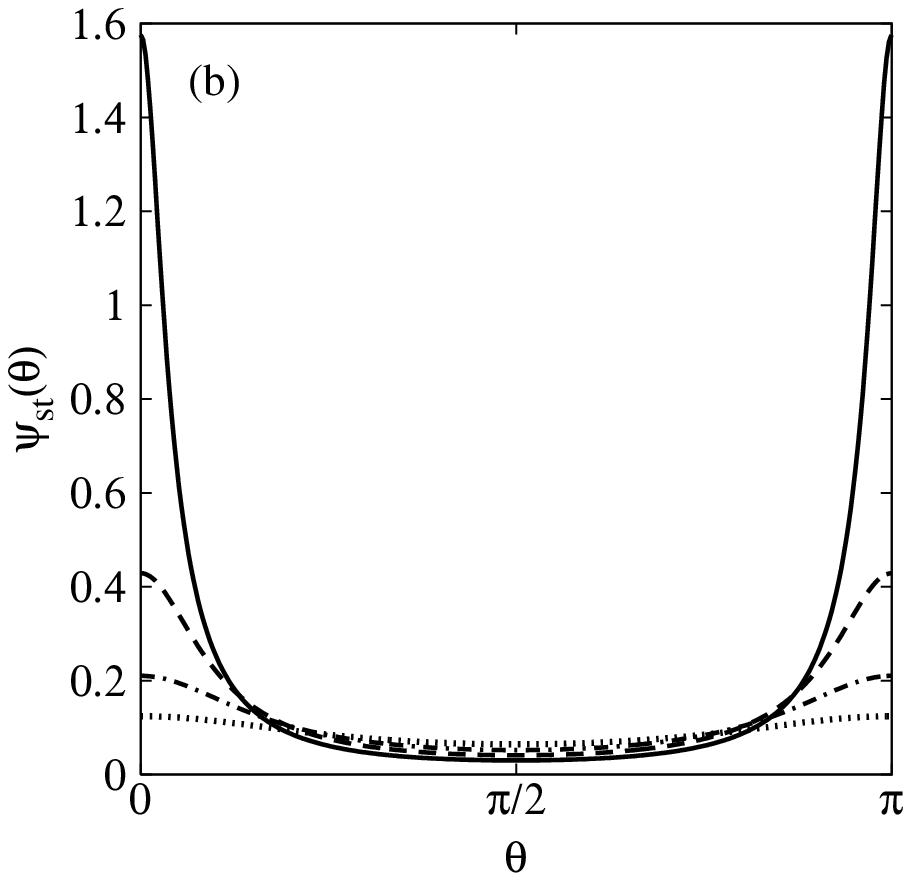}
\caption{Stationary p.d.f. of the orientation 
angle~$\vartheta$.
(\textit{a}) Rotation about~$\boldsymbol{\lambda}$: 
$b/a=4.8$, $c/a=55$, $d/a=0.85$, $\mathcal{D}'_R/a=10^{-2}$, 
and~$\gamma=-1$ (solid line), $\gamma=-0.75$
(dashed line), $\gamma=-0.5$ (dot-dashed line), $\gamma=-0.25$ (dotted line).
(\textit{b}) 
Tumbling motion between~$\boldsymbol{\lambda}$ and~$-\boldsymbol{\lambda}$: 
$b/a=4.75$, $c/a=10$, $d/a=0.9$, $\mathcal{D}'_R/a=10^{-2}$, 
and~$\gamma=1$ (solid line), $\gamma=0.75$
(dashed line), $\gamma=0.5$ (dot-dashed line), $\gamma=0.25$ (dotted line).
The normalisation coefficient~$\mathcal{N}$ has been computed numerically
according to~\eqref{eq:normalisation}.}
\label{fig:rotation+tumbling}
\end{figure}
\item\emph{Tumbling motion}.
In the following regime:
\begin{equation}
\label{eq:tumbling}
\vert\sigma\vert>\vert c\gamma\vert
\qquad \text{and} \qquad 
\gamma\sigma>0,
\end{equation}
${\psi}_{\mathrm{st}}$ has three extrema:
a minimum in~$\upi/2$ and two maxima in~$0$ and in~$\upi$
(figure~\ref{fig:rotation+tumbling}).
The particle tumbles between the direction parallel to~$\boldsymbol{\lambda}$
and that antiparallel to~$\boldsymbol{\lambda}$. 
The probability of the orientation angle~$\vartheta$ being in the neighbourhood
of~$0$ or~$\upi$
depends on the anisotropy degree of the flow and
on the shape of the particle (figure~\ref{fig:rotation+tumbling}).
\item\emph{Preferential alignment with a direction oblique to the
axis of symmetry of the flow}.
If
\begin{equation}
\vert\sigma\vert<\vert c\gamma\vert
\qquad \text{and} \qquad
c>0 ,
\end{equation}
then ${\psi}_{\mathrm{st}}$ has
three minima in~$0,\upi/2,\upi$ and two maxima
in~$\vartheta_\star$ and~$\upi-\vartheta_\star$,
where~$0<\vartheta_\star<\upi/2$ is such that
\begin{equation}
\label{eq:theta-star}
\sin\vartheta_\star
=\sqrt{\dfrac{1}{2}\left(1-\dfrac{\sigma}{c\gamma}\right)}.
\end{equation}
The particle therefore spends most of the time at an angle~$\vartheta_\star$
(or~$\upi-\vartheta_\star$) with respect to~$\boldsymbol{\lambda}$
(figure~\ref{fig:preferential+combination}).
\begin{figure}
\setlength{\unitlength}{\textwidth}
\includegraphics[width=0.495\textwidth]{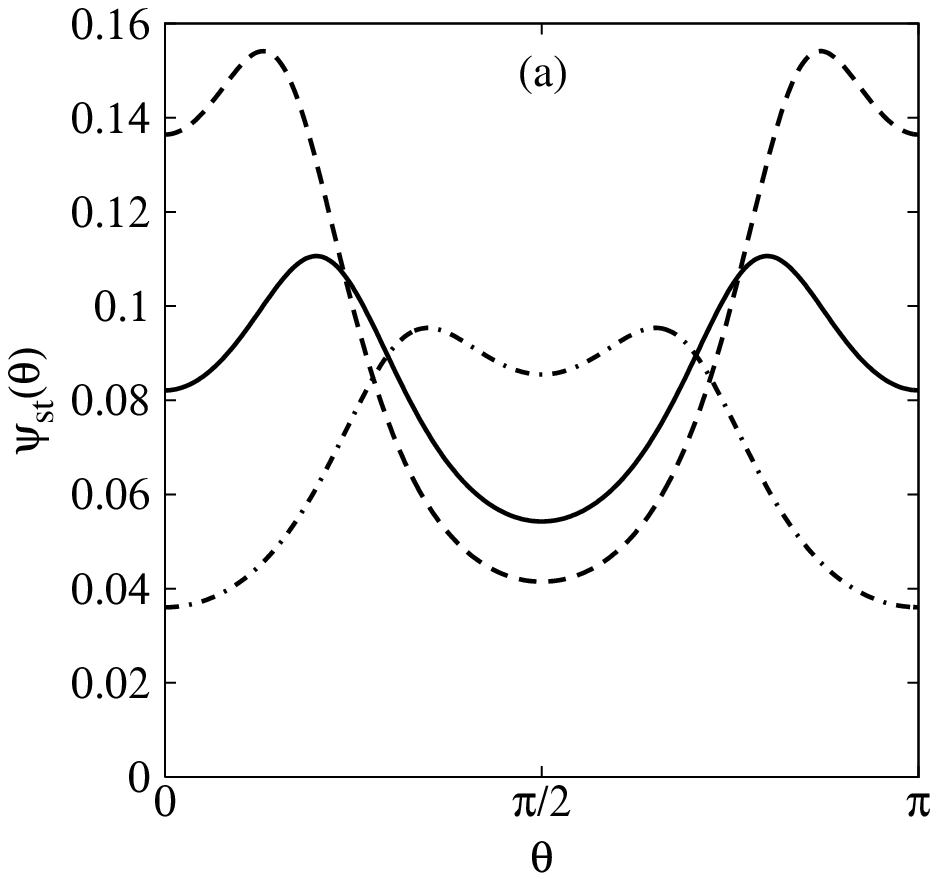}
\hfill
\includegraphics[width=0.495\textwidth]{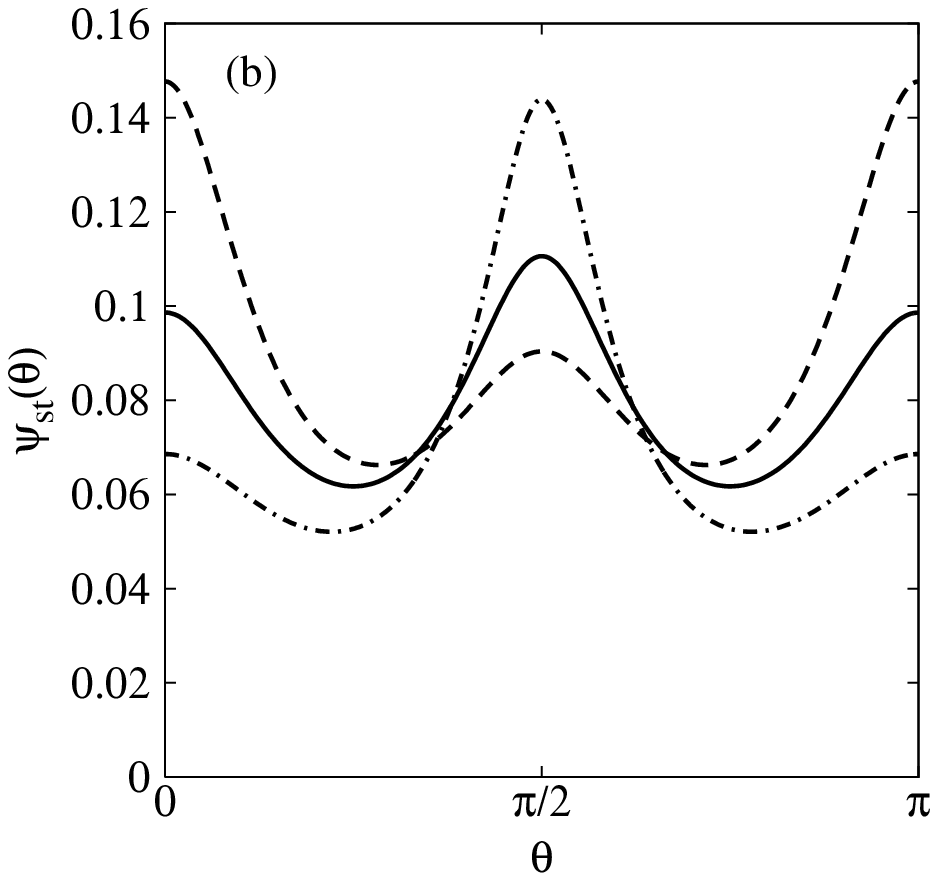}
\caption{Stationary p.d.f. of the orientation 
angle~$\vartheta$.
(\textit{a}) Preferential alignment with a direction oblique 
to~$\boldsymbol{\lambda}$: $\gamma=1$, $\mathcal{D}'_R/a=10^{-2}$,
$b/a=-6$, $d/a=-0.9$, and~$c/a=55$ (dashed line), $c/a=25$
(solid line), $c/a=11$ (dot-dashed line).
(\textit{b}) Combination of rotation and tumbling: 
$\gamma=1$, $\mathcal{D}_R'/a=10^{-2}$, $c/a=-10$, $d/a=1$,
and~$b/a=1.1$ (dot-dashed line), $b/a=2$ (solid line), $b/a=3$ (dashed line).
The p.d.f. has been normalised numerically
according to~\eqref{eq:normalisation}.}
\label{fig:preferential+combination}
\end{figure}
\item \emph{Combination of rotation and tumbling}.
For
\begin{equation}
\vert\sigma\vert<\vert c\gamma\vert
\qquad \text{and} \qquad 
c<0,
\end{equation}
the function~${\psi}_{\mathrm{st}}$ has
three maxima in~$0,\upi/2,\upi$ and two minima in~$\vartheta_\star$
and~$\upi-\vartheta_\star$ with~$\vartheta_\star$
defined in~\eqref{eq:theta-star}.
In this regime, the particle preferentially lies either
in the plane perpendicular to~$\boldsymbol{\lambda}$, in the direction
parallel to~$\boldsymbol{\lambda}$, or in the direction antiparallel to it.
\end{enumerate}

\begin{figure}
\centering
\includegraphics[width=0.495\textwidth]{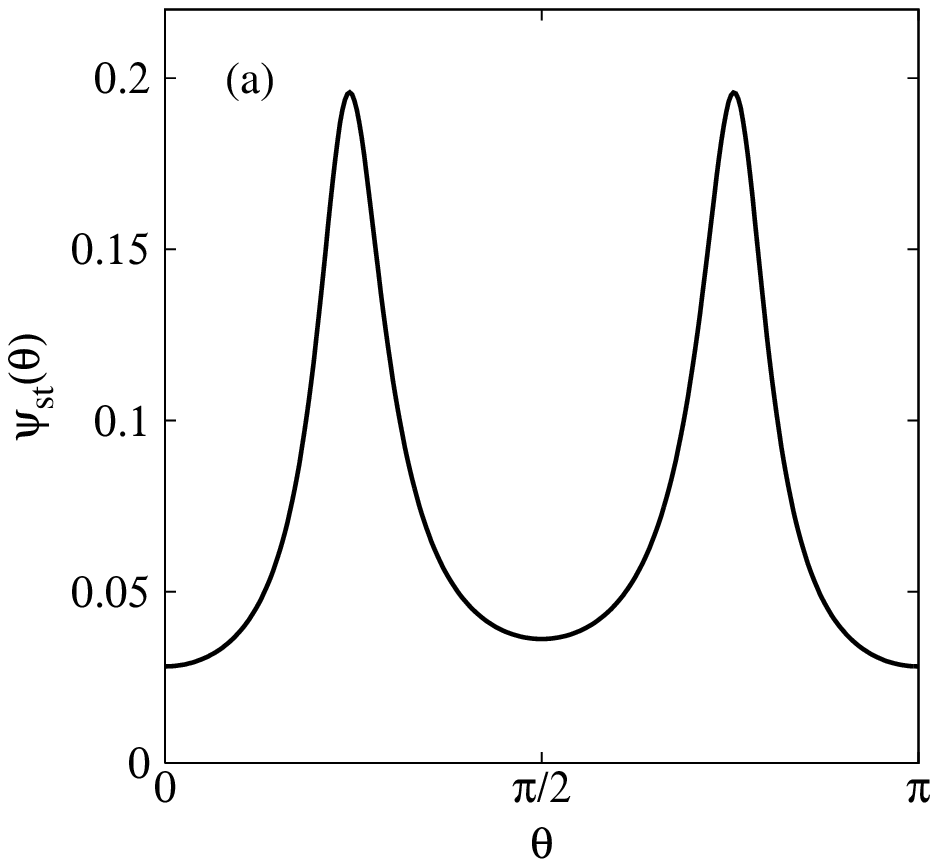}
\hfill
\includegraphics[width=0.495\textwidth]{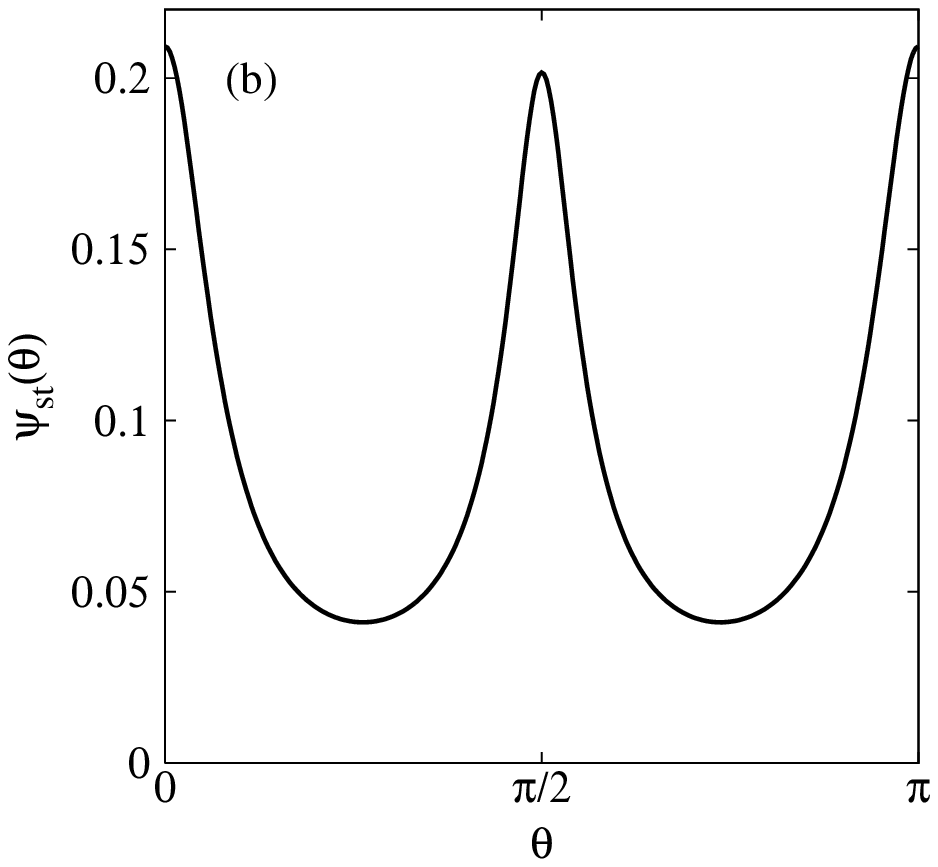}
\caption{Stationary p.d.f. of the orientation 
angle~$\vartheta$ for~$\vert\gamma\vert>1$.
(\textit{a}) Preferential alignment with a direction oblique 
to~$\boldsymbol{\lambda}$: $\gamma=5$, $\mathcal{D}'_R/a=10^{-2}$,
$b/a=-6$, $c/a=25$, $d/a=-0.9$.
(\textit{b}) Combination of rotation and tumbling: 
$\gamma=5$, $\mathcal{D}_R'/a=10^{-2}$, $b/a=2$,
$c/a=-10$, $d/a=1$.}
\label{fig:gamma>1}
\end{figure}

Naturally, the coefficient~$a$ controlling the intensity of the 
isotropic component of the flow does not play any role in the above
classification; the dynamical regime 
is selected by~$b$, $c$, $d$, and by the
shape coefficient~$\gamma$.
Whereas sufficiently elongated or flattened spheroids can be
strongly aligned in regimes (i) and~(ii), the ability of 
the flow
to orient particles is weaker in regimes~(iii) and~(iv). 
In these regimes, stronger alignment can be obtained for~$\vert\gamma\vert>1$
(figure~\ref{fig:gamma>1}). Nevertheless, \citet{B62} observed that 
particles with~$\vert\gamma\vert>1$ may be unrealistic,
albeit conceivable from a purely geometrical
point of view.

The form of~${\psi}_{\mathrm{st}}(\vartheta)$ simplifies considerably
if~$\langle\upartial_1v_1(\boldsymbol{x},t+\tau)
\upartial_1v_1(\boldsymbol{x},t)\rangle=
\langle\upartial_3v_3(\boldsymbol{x},t+\tau)
\upartial_3v_3(\boldsymbol{x},t)\rangle$, i.e. if~$d=0$.
In this case, $\mu_2=3\mu_4/4$ and hence
${\psi}_{\mathrm{st}}(\vartheta)=\mathcal{N}
[\mathcal{A}_{\vartheta\vartheta}(\vartheta)]^{-3/4}$;
furthermore, $\sigma$ does not depend on~$\gamma$.
It then follows from~\eqref{eq:rotation} and~\eqref{eq:tumbling}
that if the particles with shape coefficient~$\widehat{\gamma}$ 
rotate in the plane
orthogonal to~$\boldsymbol{\lambda}$ (resp. tumble),  the particles
with shape coefficient~$-\widehat{\gamma}$ tumble (resp. rotate
in the plane orthogonal to~$\boldsymbol{\lambda}$).
By contrast, regimes~(iii) and~(iv)
are independent on the sign of~$\gamma$, i.e. on whether the particle
is elongated or flattened,
although~$\vartheta_\star$ changes when the sign of~$\gamma$ changes.
Moreover, if the particles with shape coefficient~$\widehat{\gamma}$
rotate about~$\boldsymbol{\lambda}$ (resp. tumbles), then
all particles with~$\vert\gamma\vert<\vert\widehat{\gamma}\vert$ 
and $\text{sgn}(\gamma)=\text{sgn}(\widehat{\gamma})$ rotate
about~$\boldsymbol{\lambda}$ (resp. tumble). 
Similarly, if the particles with shape coefficient~$\widehat{\gamma}$
are in regime~(iii) (resp. in regime~(iv)), then
all particles with~$\vert\gamma\vert>\vert\widehat{\gamma}\vert$ 
are in regime~(iii) (resp. in regime~(iv)).
These properties, however, do not generally hold true if~$d$ is nonzero.
For instance, if~$b/a=-4.4$, $c/a=0.5$, $d/a=1.5$, then
particles tumble for~$1\geqslant\gamma>0.8$, they have a preferential
orientation for~$0.8>\gamma>0.4$, they rotate for~$0.4>\gamma>0$, and
they tumble again for~$0>\gamma\geqslant -1$.

\section{Conclusions}
\label{sec:conclusions}
Axisymmetric turbulence arises as 
one of the
simplest frameworks in which to study the orientation dynamics of
non-spherical particles.
On the assumptions of Gaussianity and short correlation in time,
it was shown analytically that the dynamics of a non-spherical particle
immersed in a random axisymmetric flow exhibits four regimes:
rotation around the axis of symmetry of the flow, tumbling,
combination of rotation and tumbling, and preferential
alignment with a direction oblique to the axis of symmetry of the flow.
The regime is selected by the form of the anisotropic component
of the flow and by the geometrical shape of the particle.
If the flow is weakly anisotropic or if the particle is almost spherical,
the mathematical description of~$\psi_{\mathrm{st}}(\vartheta)$
in terms of minima and maxima remains formally valid, but
the above physical classification loses its meaning, since
$\psi_{\mathrm{st}}(\vartheta)$ does not differ appreciably from the
uniform distribution.

\begin{figure}
\centering
\includegraphics[height=0.495\textwidth]{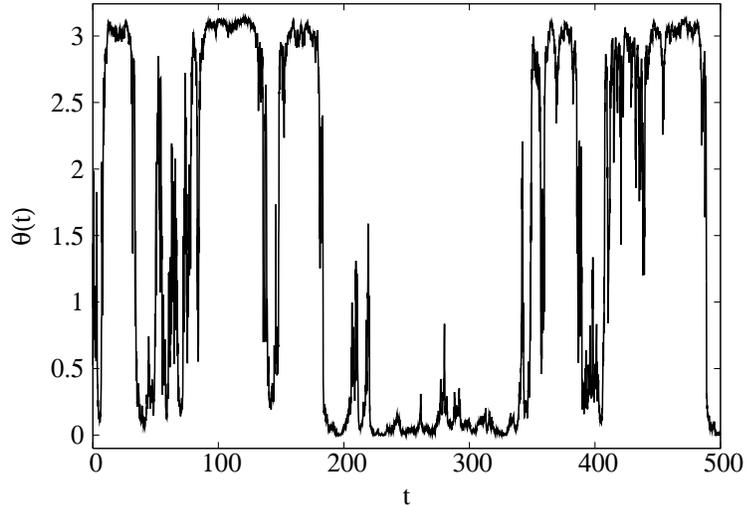}
\caption{A typical time evolution of~$\vartheta(t)$ in the tumbling regime
($\gamma=1$, $b/a=4.75$, $c/a=10$, $d/a=0.9$, $\mathcal{D}'_R/a=10^{-2}$);
$\vartheta(t)$ has been computed by numerically integrating~\eqref{eq:theta}.}
\label{fig:tumbling}
\end{figure}

The tumbling motion of a non-spherical particle
in the axisymmetric random flow differs from that of
a rod immersed in the flow resulting
from the superposition of a uniform shear and of a
short-correlated isotropic random component \citep{PT05,T07}.
In the presence of a strong mean shear, the tumbling dynamics of a rod consists of
aperiodic transitions between two unstable states: the one aligned
with the direction of the shear and that anti-aligned with it.
When a fluctuation takes the rod away from the aligned or
anti-aligned state and moves it into the unstable region of the flow
the mean shear makes the rod flip.
By contrast, in the axisymmetric case,
large deviations of the orientation of the particle from
the axis of symmetry of the flow do not necessarily result into sudden flips
of the particle (figure~\ref{fig:tumbling}). 
Simply, the orientation vector of the particle
fluctuates randomly, but the orientations aligned and anti-aligned
with the axis of symmetry of the flow are much more probable than the
other orientations. Thus, there can be excursions of~$\vartheta(t)$
from~$\vartheta\approx 0$ to~$\vartheta\approx\pi/2$ followed by a return
to~$\vartheta\approx 0$ (figure~\ref{fig:tumbling}). 
This behaviour would not be possible in the presence
of a strong mean shear.

The axisymmetric random flow and the laminar uniform shear differ in
the dependence of the orientation dynamics on the geometrical shape 
of particles.
In a uniform shear flow, 
the motion of a rod or of a disc represents a degenerate
case of the dynamics of non-spherical particles \citep{J22}.
Moreover, the dynamics is qualitatively different for~$\vert\gamma\vert<1$
and~$\vert\gamma\vert>1$ \citep{B62}.
In the axisymmetric random flow,
the dynamics of particles changes smoothly as a function of 
the shape coefficient.

Finally, it is worth remarking that the 
function~$\chi$ defined in~\eqref{eq:chi}
also determines the probability distribution of Jeffery's orbits 
in the presence of weak Brownian fluctuations \citep{LH71}.
There does not seem to be, however, a simple relation between the
orientation dynamics in the uniform shear and that in the axisymmetric
random flow.

\acknowledgements
The author is grateful to J\'er\'emie Bec, 
Antonio Celani, Fran\c{c}ois Delarue,
Prasad Perlekar, Raymond Shaw, and Michael Wilkinson for useful suggestions.
This work was supported in part by the French National Research Agency (ANR)
under grant~BLAN07-1\_192604 ``Dynamique et statistique des particules en
turbulence'' and by the EU COST Action~MP0806 ``Particles in turbulence''.

\appendix

\section{Rotary diffusion}
\label{app:A}
Rotary diffusion is introduced in the dynamics of~$\boldsymbol{N}(t)$
by assuming that its effect is to produce isotropic Brownian fluctuations
of the direction of~$\boldsymbol{N}(t)$, but not of its length.
Mathematically, this is obtained by adding a Laplacian term to the
equation for~$f(\boldsymbol{n};t)$ which acts only on the orientation 
of~$\boldsymbol{n}$. 
For a deterministic flow, $f(\boldsymbol{n};t)$ thus
satisfies \citep[e.g.][]{LH71,HL72,B74,BHAC77,DE86}:
\begin{equation}
\label{eq:FP}
\upartial_t f=  
-\widetilde{\bnabla}\bcdot[\boldsymbol{w}(\boldsymbol{n},t)f]
+\dfrac{\mathcal{D}_R}{2}{\widetilde{\nabla}}^2 f,
\end{equation}
where
\begin{equation}
\widetilde{\bnabla}\equiv{\mathsfbi{\Sigma}}(\boldsymbol{n})
\bcdot\bnabla_{\boldsymbol{n}}
\end{equation}
with~${\mathsfbi{\Sigma}}(\boldsymbol{n})
={\mathsfbi{I}}-\boldsymbol{n}\boldsymbol{n}/\vert\boldsymbol{n}\vert^2$
and~$\bnabla_{\boldsymbol{n}}=(\upartial/\upartial{n_1},\upartial/\upartial{n_2},
\upartial/\upartial{n_3})$. The differential operator~$\widetilde{\bnabla}$ 
is the angular part of the gradient
or, equivalently, the restriction of the gradient to the sphere of 
radius~$\vert\boldsymbol{n}\vert$.
In~\eqref{eq:FP},
$\widetilde{\nabla}^2\equiv\widetilde{\bnabla}\bcdot\widetilde{\bnabla}$
and $\boldsymbol{w}(\boldsymbol{n},t)=
\boldsymbol{\kappa}(t)\bcdot
\boldsymbol{n}-[\boldsymbol{\kappa}(t)
\boldsymbol{:}\boldsymbol{n}\boldsymbol{n}]\boldsymbol{n}
/\vert\boldsymbol{n}\vert^2$.

The matrix~${\mathsfbi{\Sigma}}$ satisfies: 
\begin{equation}
\label{eq:properties-sigma}
{\mathsfbi{\Sigma}}={\mathsfbi{\Sigma}}^\mathrm{T}=
{\mathsfbi{\Sigma}}^2, \qquad
\boldsymbol{n}\times[\bnabla_{\boldsymbol{n}}
\bcdot{\mathsfbi{\Sigma}}(\boldsymbol{n})]
=\boldsymbol{0}, \qquad
{\mathsfbi{\Sigma}}(\boldsymbol{n})
\bcdot{\boldsymbol{n}}=\boldsymbol{0}.
\end{equation}
By using properties~\eqref{eq:properties-sigma} and~$\boldsymbol{w}
(\boldsymbol{n},t)\bcdot\boldsymbol{n}=0$, it is possible to
rewrite~\eqref{eq:FP} as follows:
\begin{equation}
\label{eq:FP-n}
\dfrac{\upartial f}{\upartial t}=
-\dfrac{\upartial}{\upartial n_i}[w_i(\boldsymbol{n},t)f]
+\dfrac{\mathcal{D}_R}{2}
\dfrac{\upartial}{\upartial n_i}\varSigma_{ik}(\boldsymbol{n})
\dfrac{\upartial}{\upartial n_j}\varSigma_{jk}(\boldsymbol{n}) f.
\end{equation}
Equation~\eqref{eq:N} is the Stratonovich stochastic differential
equation associated with~\eqref{eq:FP-n}.

It is worth noting that
given that the orientation vector has unit length, 
$f(\boldsymbol{n};t)$ must
take the form~$f(\boldsymbol{n};t)=F(\boldsymbol{n};t)\delta(\vert
\boldsymbol{n}\vert-1)$ with~$\upartial F/\upartial n=0$.
It would therefore be more natural to consider the evolution equation 
for~$F(\boldsymbol{n};t)$ instead of that for~$f(\boldsymbol{n};t)$;
this is indeed the 
usual approach in the literature \citep{LH71,HL72,B74,BHAC77,DE86}.
In the present context, however, it is easier to
formulate the problem in~$\mathbb{R}^3$ and to move to angular
variables afterwards. 

\section{Positive semi-definiteness of the covariance tensor}
\label{app:B}
The fourth-order tensor~$\mathsfbi{\Gamma}$ is the
covariance of a Gaussian second-order tensor, and must therefore
be positive semi-definite, i.e.
\begin{equation}
\label{eq:semi-definite}
\sum_{1\leqslant i,j,p,q\leqslant 3} U_{ij}\varGamma_{ijpq}
U_{pq}\geqslant 0
\end{equation}
for all second-order tensors~$\mathsfbi{U}$.
Inequality~\eqref{eq:semi-definite} can be reinterpreted within
the theory of positive semi-definite second-order tensors \citep[e.g.][]{M08}.
Consider an invertible map $\ell$
which assigns to each pair of 
indices~$(i,j)$, $1\leqslant i,j\leqslant 3$, a single index~$\ell(i,j)$
ranging from~$1$ to~$9$. By means of the map~$\ell$, $\mathsfbi{U}$ 
can be regarded  as a 9-dimensional vector; likewise, $\mathsfbi{\Gamma}$ 
can be regarded as a  symmetric
$9\times 9$ second-order tensor, whose symmetry
follows from~$\varGamma_{ijpq}=\varGamma_{pqij}$.
Accordingly, \eqref{eq:semi-definite}
can be rewritten in the following form:
\begin{equation}
\label{eq:semi-definite-2}
\sum_{1\leqslant \ell(i,j),\ell(p,q)\leqslant 9} 
U_{\ell(i,j)}\varGamma_{\ell(i,j)\ell(p,q)}U_{\ell(p,q)}\geqslant 0.
\end{equation}
Inequality~\eqref{eq:semi-definite-2} is the definition of
positive semi-definiteness for second-order tensors.
The theory of such tensors says that a necessary and sufficient condition
for an Hermitian second-order tensor to be positive semi-definite
is that all the principal minors of the tensor are
non-negative \citep[][p.~307]{G77}. 
When this condition is applied to~$\varGamma_{\ell(i,j)\ell(p,q)}$ 
it yields the following inequalities:
\begin{equation}
\label{eq:ineq-weak-1}
\begin{cases}
a+d\geqslant 0 & \text{if $a\ge 0$} \\
5a+d\geqslant 0 & \text{if $a<0$} 
\end{cases}
\end{equation}
and
\begin{eqnarray}
\label{eq:ineq-weak-2}
\varpi_1\equiv 4a-b+d\geqslant 0, \qquad \varpi_2\equiv 4a+b+c+6d\geqslant 0,
\\
\varpi_3\equiv 15a^2-b^2-(b-d)(c+5d)+2a(2c+13d)\geqslant 0.
\end{eqnarray}
In this paper, the above inequalities are assumed to hold strictly;
\eqref{eq:inequalities} follows from this assumption.

Some useful inequalities can be derived 
from~\eqref{eq:inequalities}. Here, $\gamma\neq 0$ (the case of a spherical
particle is indeed trivial), $\mathcal{D}'_R=0$ (it is easily seen
that a positive~$\mathcal{D}'_R$ does not alter the inequalities below),
and~$\mu_3\neq 0$ (the case~$\mu_3=0$ is treated separately in
appendix~\ref{app:D}).

Firstly, $\mathcal{A}_{\vartheta\vartheta}(\vartheta)$
and~$\mathcal{A}_{\varphi\varphi}(\vartheta)$ are positive for all
values of~$a$, $b$, $c$, $d$, and~$\gamma$ satisfying~\eqref{eq:inequalities}.
Indeed, the quantities~$2\mu_1$ and~$2\mu_1+\mu_3+\mu_4$ are
quadratic polynomials in~$\gamma$, 
their discriminant is equal to~$-\varpi_3<0$, and for~$\gamma=1$
they are equal to~$\varpi_1>0$ and to~$\varpi_2>0$, respectively.
Moreover, $2\mu_1+\mu_5$ is a quadratic polynomial in~$\gamma$ with
discriminant equal to~$-(3a+d)(5a+d)<0$ and takes the value~$4a+d>0$
for~$\gamma=1$.
Hence
\begin{equation}
\label{eq:ineq-mu}
\mu_1>0,\qquad 2\mu_1+\mu_5>0,\qquad 
2\mu_1+\mu_3+\mu_4>0
\end{equation}
for all~$a$, $b$, $c$, $d$, and~$\gamma$, and consequently
\begin{eqnarray}
&\mathcal{A}_{\vartheta\vartheta}(0)=
\mathcal{A}_{\vartheta\vartheta}(\upi)=2\mu_1>0&
\\
&\displaystyle
\lim_{\vartheta\to 0}\mathcal{A}_{\varphi\varphi}(\vartheta)=
\lim_{\vartheta\to \upi}\mathcal{A}_{\varphi\varphi}(\vartheta)=+\infty&
\end{eqnarray}
and
\begin{eqnarray}
\mathcal{A}_{\vartheta\vartheta}(\vartheta)
&\geqslant& \mu_4\sin^2(\vartheta)+(2\mu_1+\mu_3)\sin^4(\vartheta)
\geqslant (2\mu_1+\mu_3+\mu_4)\sin^4(\vartheta)>0
\\
\mathcal{A}_{\varphi\varphi}(\vartheta)&\geqslant&2\mu_1+\mu_5>0
\end{eqnarray}
for all~$0<\vartheta<\upi$.

Secondly, $\psi_{\mathrm{st}}(\vartheta)$ is bounded
for all~$a$, $b$, $c$, $d$, and~$\gamma$ and
for all~$0\leqslant\vartheta\leqslant\upi$.
For~$\varDelta>0$, this property is obvious.
For~$\varDelta<0$, note that
\begin{equation}
\label{eq:product-1}
(\sqrt{-\varDelta}+\mu_4)(\sqrt{-\varDelta}-\mu_4)=
-\mu_3 P_1(\gamma),
\end{equation}
and
\begin{equation}
\label{eq:product-2}
(\sqrt{-\varDelta}+\mu_4+2\mu_3)(\sqrt{-\varDelta}-\mu_4-2\mu_3)=
-\mu_3 P_2(\gamma),
\end{equation}
where~$P_1(\gamma)$ and~$P_2(\gamma)$ are quadratic polynomials in~$\gamma$
such that~$P_1(1)=4\varpi_1>0$ and~$P_2(1)=4\varpi_2>0$. 
Furthermore, the discriminants of~$P_1(\gamma)$
and~$P_2(\gamma)$ are equal to~$-16\varpi_3<0$.
Therefore~$P_1(\gamma)$ and~$P_2(\gamma)$ are positive
for all~$\gamma$, and the 
products on the left-hand-sides of~\eqref{eq:product-1}
and~\eqref{eq:product-2} have the same sign as~$-\mu_3$.

Also observe that
\begin{equation}
\label{eq:product-3}
(\sqrt{-\varDelta}+\mu_4+2\mu_3)(\sqrt{-\varDelta}-\mu_4)=
-\mu_3 \left[P_3(\gamma)-2\sqrt{-\varDelta}\right],
\end{equation}
where~$P_3(\gamma)$ is a quadratic polynomial satisfying:
$P_3^2(\gamma)+4\varDelta=P_1(\gamma)P_2(\gamma)>0$.
Hence~$P_3(\gamma)-2\sqrt{-\varDelta}\neq 0$ for all~$\gamma$.
Moreover, for~$\gamma=0$, $P_3(0)-2\sqrt{-\varDelta}=
P_3(0)=\varpi_1+\varpi_2+2(a+d)>0$. As a result
$P_3(\gamma)-2\sqrt{-\varDelta}>0$ for all~$\gamma$.
The left-hand-side of~\eqref{eq:product-3} therefore has the same
sign as~$-\mu_3$.

Three cases should now be distinguished:
\begin{enumerate}
\item $\mu_3<0$: in this case, $\sqrt{-\varDelta}\,\pm\mu_4>0$
and~$\sqrt{-\varDelta}-\mu_4-2\mu_3>0$ (remember that~$\varDelta=
8\mu_1\mu_3-\mu_4^2$ with~$\mu_1>0$), 
whence~$\sqrt{-\varDelta}+\mu_4+2\mu_3>0$ (see~\eqref{eq:product-2}). 
Therefore
\begin{subeqnarray}
\label{eq:case-1}
\sqrt{-\varDelta}+\mu_4+2\mu_3\sin^2(\vartheta)&\geqslant&
\sqrt{-\varDelta}+\mu_4+2\mu_3>0,
\\
\sqrt{-\varDelta}-\mu_4-2\mu_3\sin^2(\vartheta)&\geqslant&\sqrt{-\varDelta}-\mu_4>0.
\end{subeqnarray}
\item $\mu_3>0$ and~$\mu_4>0$: then~$\sqrt{-\varDelta}+\mu_4>0$
and consequently~$\sqrt{-\varDelta}-\mu_4<0$ (see~\eqref{eq:product-1}). Hence
\begin{subeqnarray}
\label{eq:case-2}
\sqrt{-\varDelta}+\mu_4+2\mu_3\sin^2(\vartheta)&\geqslant&
\sqrt{-\varDelta}+\mu_4>0,
\\
\sqrt{-\varDelta}-\mu_4-2\mu_3\sin^2(\vartheta)&\leqslant&\sqrt{-\varDelta}-\mu_4<0.
\end{subeqnarray}
\item $\mu_3>0$ and~$\mu_4<0$: for these values of the parameters,
\eqref{eq:product-2} and~\eqref{eq:product-3} yield the following relations:
\begin{subeqnarray}
\label{eq:case-3}
\sqrt{-\varDelta}+\mu_4+2\mu_3\sin^2(\vartheta)&\leqslant&
\sqrt{-\varDelta}+\mu_4+2\mu_3<0,
\\
\sqrt{-\varDelta}-\mu_4-2\mu_3\sin^2(\vartheta)&\geqslant
&\sqrt{-\varDelta}-\mu_4-2\mu_3>0.
\end{subeqnarray}
Inequalities~\eqref{eq:case-1}, \eqref{eq:case-2}, and~\eqref{eq:case-3} 
guarantee that if~$\varDelta<0$, the function~$\chi(\vartheta)$ 
is bounded for all~$a$, $b$, $c$, $d$, and~$\gamma$.
\end{enumerate}

For~$\varDelta=8\mu_1\mu_3-\mu_4^2=0$, $\mu_4$ cannot be zero
since~$\mu_1>0$ and~$\mu_3\neq 0$. Moreover, $\mu_3$ and hence~$c$ must 
be positive, in that~$\mu_1>0$. Therefore, for~$\mu_4>0$ the 
function~$\chi(\vartheta)$ is bounded. The case~$\mu_4<0$ requires a more
detailed analysis. $\varDelta$ can be rewritten thus: $\varDelta=
c(6a+5d)\gamma^2(\gamma^2-\rho)$ with
\begin{eqnarray}
\rho=\frac{(2b+c+5d)^2-c(10a+c+9d)}{c(6a+5d)}.
\end{eqnarray}
(Note that, for~$\gamma=0$, $\mu_1=(10a+c+9d)/8$ and hence~$10a+c+9d>0$
for all~$a$, $c$, $d$; this is consistent with the positivity of
the variance of the components of the vorticity --- see~\eqref{eq:vorticity-1}.)
Provided that~$\rho>0$, $\varDelta$ vanishes for~$\gamma=\pm\gamma_\star$
with~$\gamma_\star=\sqrt{\rho}$ (the case~$\gamma=0$ is not 
considered here). Now note that both~$\mu_4$ and~$2\mu_3+\mu_4$ are quadratic polynomials
in~$\gamma$. If~$\gamma_{\star\star}\equiv 1+(2b+5d)/c$ is positive, then
\begin{equation}
\begin{cases}
\mu_4>0 & \text{if $\gamma\in(0,\gamma_{\star\star})$}
\\
\mu_4< 0 & \text{if $\gamma\notin[0,\gamma_{\star\star}]$}
\end{cases}
\qquad \text{and} \qquad
\begin{cases}
2\mu_3+\mu_4<0 & \text{if $\gamma\in(-\gamma_{\star\star},0)$}
\\
2\mu_3+\mu_4> 0 & \text{if $\gamma\notin[-\gamma_{\star\star},0]$}.
\end{cases}
\end{equation}
If~$\gamma_{\star\star}<0$, then
\begin{equation}
\begin{cases}
\mu_4>0 & \text{if $\gamma\in(-\gamma_{\star\star},0)$}
\\
\mu_4< 0 & \text{if $\gamma\notin[-\gamma_{\star\star},0]$}
\end{cases}
\qquad \text{and} \qquad
\begin{cases}
2\mu_3+\mu_4<0 & \text{if $\gamma\in(0,\gamma_{\star\star})$}
\\
2\mu_3+\mu_4> 0 & \text{if $\gamma\notin[0,\gamma_{\star\star}]$}.
\end{cases}
\end{equation}
Hence
\begin{equation}
\label{eq:ineq-delta-0}
\mu_4(2\mu_3+\mu_4)>0 \qquad \forall\; \vert\gamma\vert<\vert
\gamma_{\star\star}\vert.
\end{equation}
Let us now show that~$\vert\gamma_\star\vert<\vert\gamma_{\star\star}\vert$.
The quantity~$\gamma_{\star\star}^2-\gamma_{\star}^2$ is written:
\begin{equation}
\gamma_{\star\star}^2-\gamma_{\star}^2=\dfrac{P_4(c)}{(6a+5d)c^2},
\end{equation}
where~$P_4(c)$ is a quadratic polynomial in~$c$ whose coefficients depend
on~$a$, $b$, and~$d$ and such that~$P_4(0)=(6a+5d)(2b+5d)^2>0$
and~$\lim_{c\to\infty}P_4(c)/c^2=4\varpi_1>0$. Moreover, $P'_4(c)=0$
if and only if~$c=\widehat{c}\equiv -(2b+5d)(12a-2b+5d)/(8\varpi_1)$,
and for~$c=\widehat{c}$:
\begin{equation}
P_4(\widehat{c})=\dfrac{(2b+5d)^2[16\varpi_3+(2b+5d)^2]}{16\varpi_1}>0.
\end{equation}
Therefore, $P_4(c)>0$ for all~$c>0$ and~$\gamma_{\star\star}^2>
\gamma_\star^2$ for all~$c>0$. As a conclusion, if~$\varDelta=0$, 
then~\eqref{eq:ineq-delta-0} holds, and for~$\mu_4<0$:
\begin{equation}
\mu_4+2\mu_3\sin^2\vartheta\leqslant\mu_4+2\mu_3<0 \qquad \forall\; 0\leqslant
\vartheta\leqslant\upi.
\end{equation}
This result proves that~$\chi(\vartheta)$ is bounded also for~$\varDelta=0$
and~$\mu_4<0$.

\section{Stationary probability density function 
of the orientation angle}
\label{app:C}
Consider the functions:
\begin{equation}
\varPhi(\vartheta)=\ln\left[\dfrac{\mathcal{A}_{\vartheta\vartheta}(\vartheta)}{2}\right]-2\int_{\vartheta_0}^\vartheta\dfrac{\mathcal{B}_\vartheta(z)}%
{\mathcal{A}_{\vartheta\vartheta}(z)}\,\mathrm{d}z
\end{equation}
and
\begin{equation}
g(\vartheta)=2\int_{\vartheta_0}^\vartheta\dfrac{\mathrm{e}^{\varPhi(z)}}%
{\mathcal{A}_{\vartheta\vartheta}(z)}\,\mathrm{d}z.
\end{equation}
The stationary solution of~\eqref{eq:FPE-spherical} is written
\citep[see][p.~98]{R89}:
\begin{equation}
\label{eq:C3}
{\varPsi}_{\mathrm{st}}(\vartheta)=
\mathcal{N}\mathrm{e}^{-\varPhi(\vartheta)}-\mathcal{S}g(\vartheta)
\mathrm{e}^{-\varPhi(\vartheta)},
\end{equation}
where~$\mathcal{N}$ and~$\mathcal{S}$ are constants.
As~$\mathcal{A}_{\vartheta\vartheta}(\vartheta)$
and~$\mathcal{B}_{\vartheta}(\vartheta)$ are periodic, also 
$\mathrm{e}^{-\varPhi(\vartheta)}$ is periodic.
By contrast,
$g(\vartheta)$ cannot be periodic  
given that~${\mathcal{A}_{\vartheta\vartheta}(\vartheta)}> 0$ and 
hence~$g'(\vartheta)>0$.
Therefore, ${\varPsi}_{\mathrm{st}}(\vartheta)$ 
satisfies~\eqref{eq:periodic-bc} if and only
if~$\mathcal{S}=0$. Equation~\eqref{eq:Psi-stationary} then follows
from~\eqref{eq:C3} with~$\mathcal{S}=0$.

\section{The case $\mu_3=0$}
\label{app:D}
If~$\mu_3=0$ and~$\gamma\neq 0$ (i.e.~$c=0$), 
three cases should be distinguished.

For~$\mu_3=0$, $\mu_4\neq 0$, and~$\mu_2\neq 3\mu_4/2$,
the integral in~\eqref{eq:Psi-stationary} 
can be easily calculated
by means of the transformation~$y=\sin^2(\omega)$ to yield:
\begin{equation}
\label{eq:D1}
{\psi}_{\mathrm{st}}(\vartheta)=\mathcal{N}\left(2\mu_1+\mu_4\sin^2\vartheta
\right)^{\frac{\mu_2}{\mu_4}-\frac{3}{2}}.
\end{equation}
The stationary p.d.f. is bounded as a consequence of~\eqref{eq:ineq-mu}.
By examining the first derivative of the above function, it can be shown
that, depending on the value of the parameters,
${\psi}_{\mathrm{st}}$ has either three or five extrema
in the interval~$0\leqslant \vartheta\leqslant\upi$. 
Therefore,
the four regimes identified for~$\mu_3\neq 0$ also describe the
dynamics of the particle for~$\mu_3=0$.

For~$\mu_3=\mu_4=0$ (i.e.~$c=2b+5d=0$), the stationary solution
is:
\begin{equation}
\label{eq:D2}
{\psi}_{\mathrm{st}}(\vartheta)=\mathcal{N}
\exp\left[\dfrac{\mu_2}{2\mu_1}\sin^2(\vartheta)\right].
\end{equation}
For~$\mu_2/\mu_1>0$, ${\psi}_{\mathrm{st}}$ has
two minima in~0 and~$\upi$ and one maximum in~$\upi/2$, and hence the
particle rotates in the plane orthogonal to~$\boldsymbol{\lambda}$.
For~$\mu_2/\mu_1<0$, ${\psi}_{\mathrm{st}}$ has 
two maxima in~0 and~$\upi$ and one minimum in~$\upi/2$; therefore the particle
tumbles between the direction parallel to~$\boldsymbol{\lambda}$
and that antiparallel to~$\boldsymbol{\lambda}$. 

Finally, 
for~$\mu_3=0$ and~$\mu_2=3\mu_4/2$ (i.e.~$c=2b+(5+\gamma)d=0$),
\eqref{eq:D1} implies that
the stationary statistics of orientations is isotropic:
$\psi_{\mathrm{st}}(\vartheta)=(4\upi)^{-1}$.

\end{document}